\documentclass[aps,prl,longbibliography, nofootinbib, twocolumn,amsmath,amssymb,10pt]{revtex4-1}
\usepackage{graphicx}
\usepackage[normalem]{ulem}
\usepackage{soul}
\usepackage{txfonts}
\usepackage[colorlinks, linkcolor=blue]{hyperref}
\usepackage{verbatim}
\usepackage{esint}
\usepackage{amsmath}
\usepackage{mathtools}

\def \be {\begin{equation}}
\def \ee {\end{equation}}

\begin{document}

\title{The fractal dimension of the Appalachian Trail}
\author{Brian Skinner}
\affiliation{Department of Physics, The Ohio State University, Columbus, Ohio 43202, USA}

\date{\today \\
\vspace{.1in}}

\begin{abstract}

The Appalachian Trail (AT) is a 2193-mile-long hiking trail in the eastern United States. The trail has many bends and turns at different length scales, which give it a nontrivial fractal dimension. Here I use GPS data from the Appalachian Trail Conservancy to estimate the fractal dimension of the AT. I find that, at length scales between $\sim 20$\,m and $\sim 100$\,km, the trail has a well-defined ``divider dimension'' of $\approx 1.08$. This dimension can be used to estimate the true hiking distance between two points, given the distance as estimated from a map with finite spatial resolution (e.g., Google Maps).

\end{abstract}

\maketitle

A fractal is a geometrical object for which the apparent total extent depends in an unusual way on the magnification with which the object is viewed \cite{mandelbrot_fractal_1982}. As a particularly striking example, Mandelbrot pointed out in 1967 that the apparent length of the coastline of Britain depends sensitively on the length scale with which it is resolved \cite{mandelbrot_how_1967}, such that measuring with a shorter ruler produces a longer estimate for the total length of the coast. In fact, a similar issue was pointed out in 1961 by Lewis Fry Richardson, who helped to resolve a dispute between Spain and Portugal about the length of their border \cite{richardson1961problem, mccartney_how_2008}. Richardson showed that if one approximates a jagged curve, such as the one defining the border of a country, by a series of straight line segments of length $\ell$, then the apparent total contour length $L$ of the curve varies as $\ell^{D-1}$, where $D$ is a nontrivial number between $1$ and $2$. A plot of $\log L$ vs.\ $\log \ell$ is referred to as a ``Richardson plot'' \cite{mccartney_how_2008}; its slope gives $D - 1$.  This definition of $D$ is commonly referred to as the ``divider dimension'', to distinguish it from alternate measures of fractal dimension such as the ``Hausdorff dimension'' or ``box counting dimension'' \cite{dunbar_divider_1992, klotzbach_new_1998}.

\begin{figure}[h!]
\begin{center}
\includegraphics[width=0.95\columnwidth]{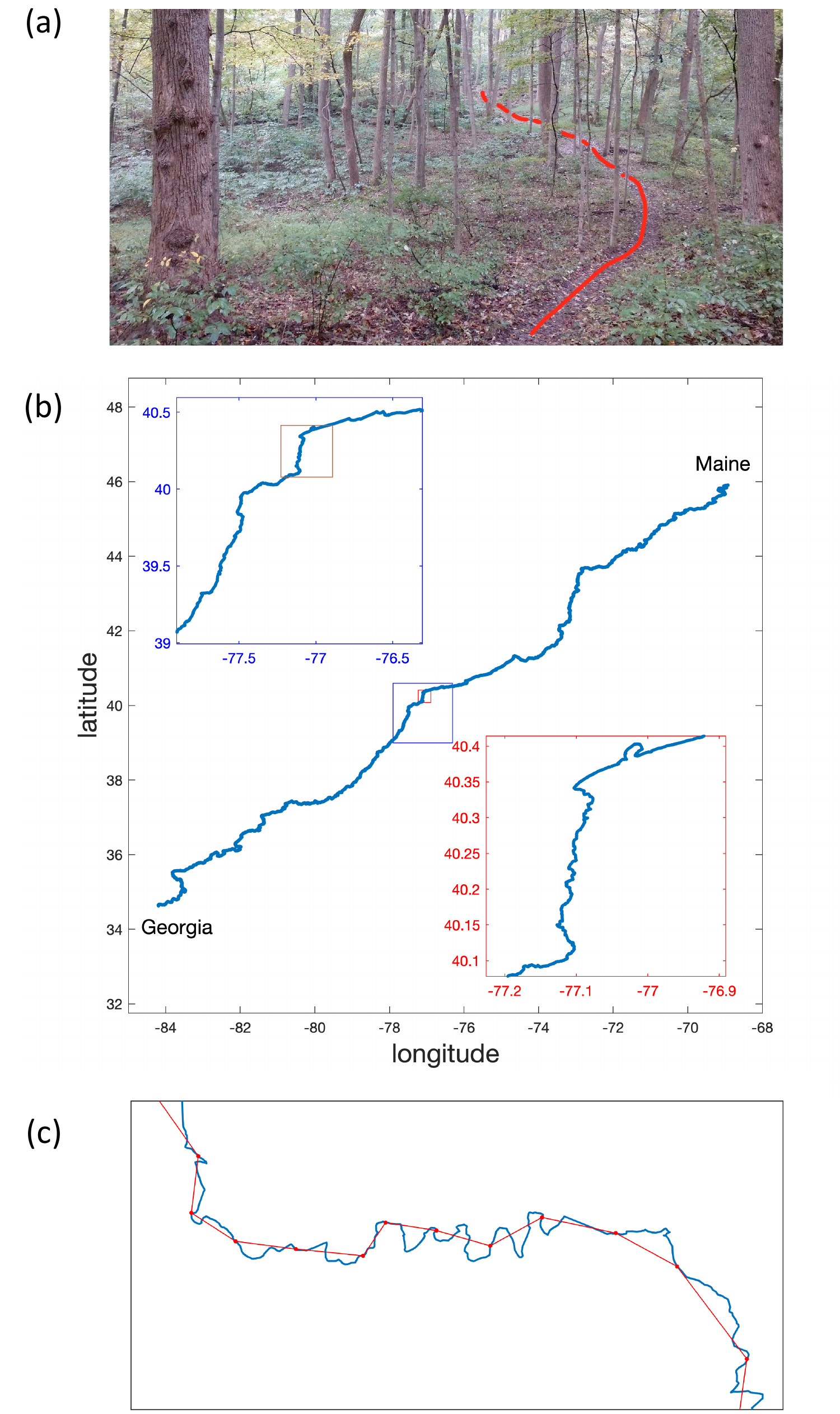}
\end{center}
\caption{(a) A typical hikers-eye view of the AT, showing meandering of the trail (red line). (b) The full AT, shown in equirectangular projection along with segments at different levels of magnification (insets). (c) A segment of the AT (thick blue line), is shown along with a shorter curve (thin red line) formed by placing one point per quarter mile of walking distance and connecting the points by straight line segments.}
\label{fig:Fig1}
\end{figure}

While underestimating the length of a country's border can provoke an international incident, underestimating the length of a hiking trail can cause trouble of its own. Inspired by a recent personal hiking trip (which had some underestimation problems), in this short article I consider the fractal dimension of the Appalachian Trail (AT). The AT runs through a forested region of the eastern United States, and is considered to be the longest hiking-only trail in the world. Its official length (as of 2020) is 2193 miles (3529 km), \cite{noauthor_interactive_nodate} but the Cartesian separation between its two endpoints is only 1114 miles (1792 km). At much shorter length scales the trail looks jagged, with an apparently self-similar structure (Fig.\ \ref{fig:Fig1}). 

Global positioning system (GPS) data for the AT is publicly available via the Appalachian Trail Conservancy (downloadable at \cite{noauthor_appalachian_nodate}). In this data set, the trail is resolved at about one point per eleven meters (there are $N = 312,036$ points in total). As shown below, this eleven meter resolution is apparently below the range of length scales over which the AT looks fractal, so that the separation between two subsequent points in the GPS data is a good estimation of the true walking distance between them. Each GPS point entails a latitude, longitude, and elevation relative to sea level. The distance between two GPS points can be calculated by first converting the GPS coordinates of each point to Cartesian coordinates:
\begin{eqnarray}
x & = & (R_e + h) \cos \phi \cos \lambda, \nonumber \\
y & = & (R_e + h) \cos \phi \sin \lambda, \nonumber \\
z & = & (R_e + h) \sin \phi, \nonumber
\end{eqnarray}
where $\phi$ is the latitude, $\lambda$ is the longitude, and $h$ is the elevation. The radius of the Earth is taken to be a constant, $R_e \approx 6371$\,km. The distance between two points $i$ and $j$ is then given by
\be 
d_{ij} = \sqrt{(x_i - x_j)^2 + (y_i - y_j)^2 + (z_i - z_j)^2}.
\ee 

In order to construct a Richardson plot, I make a variable subsampling of the GPS data, dividing the full length of the trail into an integer number $n$ of intervals. Each interval contains $(N-1)/n$ data points, and one can construct an estimated path length $L$ by summing the total straight-line distance $d_{ij}$ between all pairs of neighboring interval end points $i,j$ [as illustrated in Fig.\ \ref{fig:Fig1}(c)]. In cases where $(N-1)/n$ is not an integer, the end point of each interval is estimated by linear interpolation between the two closest GPS points. The associated segment length $\ell$ is calculated as the mean length of all $n$ segments.

\begin{figure}[htb]
\begin{center}
\includegraphics[width=0.9\columnwidth]{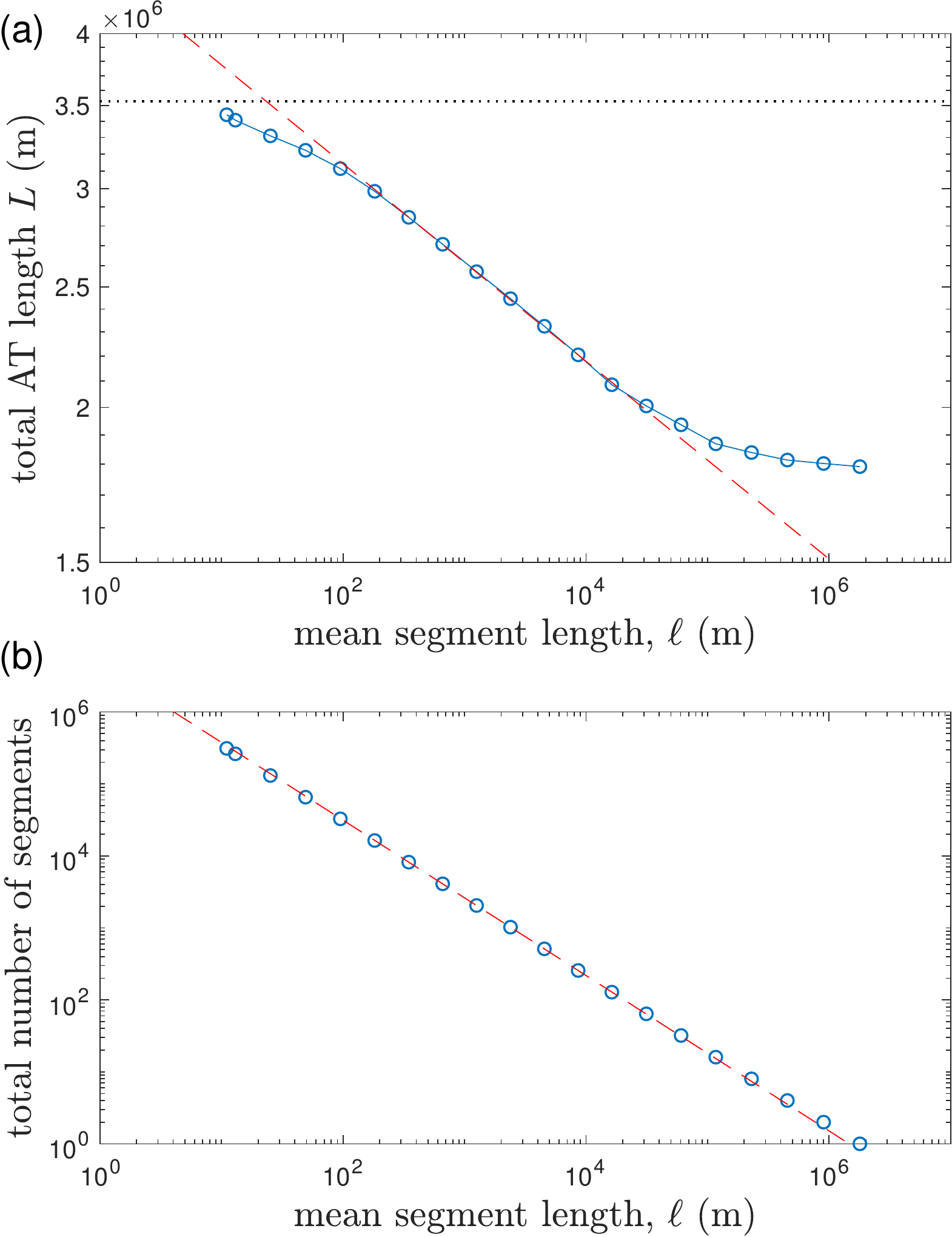}
\end{center}
\caption{(a) Richardson plot, showing the estimated total length $L$ of the AT vs.\ the mean segment length $\ell$ in double-logarithmic scale. Fitting to a power-law dependence gives $D-1$ (dashed red line). The dotted line shows the official estimate of the trail length from the Appalachian Trail Conservancy, and the dashed line is a fit to $L \propto 1/\ell^{D-1}$. (b) One can equivalently estimate $D$ by plotting the total number of segments $n$ as a function of the segment length $\ell$.}
\label{fig:Richardson}
\end{figure}

Figure \ref{fig:Richardson}(a) shows the results of this analysis. As the AT is divided into an increasingly large number $n$ of segments, the estimated total trail length $L$ increases, as in the famous ``coastline of Britain'' problem. When the segment length $\ell$ is longer than $\ell_0 \approx 23$\,m and shorter than $\approx 100$\,km, the dependence $L$ vs.\ $\ell$ is well described by a power law, $L \propto 1/\ell^{D-1}$, with $D = 1.080 \pm 0.005$. The length scale $\ell_0$ represents the typical length of a single bend in the trail (in the language of polymers, $\ell_0$ would be called the ``persistence length''). The longer length scale $\approx 100$\,km is presumably associated with large-scale planning of the trail, or with the size of topographical features, so that on length scales longer than about $100$\,km the trail looks straighter than one would expect from its fractal dimension.

An equivalent method for extracting the divider dimension $D$ is to plot the total number of segments $n$ against the segment length $\ell$, and extract the power law dependence $n \propto 1/\ell^D$. Such a plot is shown in Fig.\ \ref{fig:Richardson}(b), with the same value of $D = 1.08$.

One implication of Fig.\ \ref{fig:Richardson}(a) is that, if one is estimating the length of a hike along the AT, then an estimate made using a spatial resolution much longer than $\ell_0$ will significantly underestimate the hiking distance. In particular, if the trail is approximated by straight line segments of length $\ell_r$, then the true hiking distance $L_\text{true}$ is related to the estimated distance $L_\text{est}$ by
\be 
L_\text{true} = L_\text{est} \times \left( \frac{\ell_r}{\ell_0} \right)^{D-1}.
\ee 
For the AT, $\ell_0 \approx 23$\,m.  So, for example, if the map one is using has a resolution of about one point per quarter mile ($400$\,m), then the true hiking distance will be longer than the map-estimated hiking distance by about 25\%.%
\footnote{This hypothetical example was apparently realized by me during a recent hiking trip on the AT through Virginia, using distance estimates from Google Maps.}


\bibliography{ATfractal.bib}

\end{document}